\documentclass[a4paper,11pt,showpacs]{article}

\usepackage{jheppub} 
\usepackage{lineno}
\usepackage{tensor}

\arxivnumber{2607.01419} 

\title{Vacuum Cherenkov radiation in supercritical magnetic fields.}

\author[1]{Daniel Gálvez-García\note{Corresponding author.}}
\author{and Nora Bretón}
\affiliation{Physics Department, Cinvestav,\\Apartado Postal 14--740, C.P. 07360, Mexico City, Mexico. }

\emailAdd{daniel.galvez@cinvestav.mx}

\abstract{
In the presence of very intense electromagnetic fields, the refractive index of vacuum is modified such that light velocity is less than $c$ and ultrarelativistic charged particles can be faster than light  and can induce Cherenkov radiation. We present the comparison of the Cherenkov radiation produced by
the Euler-Heisenberg theory for critical and supercritical magnetic fields. We also make the comparison between the Cherenkov  and the synchrotron radiation produced by the charged particles. 
}

\begin{document}
\maketitle
\flushbottom



 \section{Introduction.}

In the presence of very intense magnetic fields that are constant and spatially uniform, 
the vacuum becomes polarized with a refractive index greater than unity.  Whenever we refer to vacuum we mean ``a region of space with no sources of electromagnetic field", although a magnetic or electric field generated somewhere else can (and will be) present.
In those regions charged particles can travel faster than light, therefore the conditions for the production of the Cherenkov radiation are met and such radiation can be measured. \\

Several phenomena of light propagating in very intense electromagnetic fields are described in a phenomenological way by classical theories of nonlinear electrodynamics. Such phenomena include slowdown of photon velocities, vacuum birefringence and modification of lightlike geodesics. These effects vary from one NLED theory to another and are, therefore, the most promising candidates to test the validity of each model. For instance, Born-Infeld theory \cite{BI}, \cite{Aiello2007} was derived with the aim of avoiding the divergences of a point-like charged particle's field and energy, it includes an \textit{ad hoc} upper bound for the electric field and does not predict birefringence. On the other hand, the Euler-Heisenberg (EH) theory \cite{EH}, \cite{Ruffini2013} includes corrections to Maxwell electrodynamics due to second-order terms in QED calculations, usually attributed to vacuum polarization effects when the external field approaches Schwinger's critical field, $E_{\rm cr}= cB_{\rm cr}= m_{e}^2c^3/(e \hbar) \sim 10^{18} V/m$ \cite{Schwinger1951}.\\

The form of Euler-Heisenberg lagrangian depends upon the field intensity we are dealing with. For fields below Schwinger's one shall use the \textit{weak} Euler - Heisenberg lagrangian (wEH) (\ref{eq:wEH}), while for field of order or greater than Schwinger's the correct theory would be the \textit{strong} Euler - Heinseberg lagrangian (sEH) (\ref{eq:sEH}) \cite{Karbstein2019}. In terms of the electromagnetic invariants $S= -F_{\mu \nu} F^{\mu \nu}/4$ and $P= -F_{\mu \nu} \tilde{F}^{\mu \nu}/4$,

\begin{equation}
    \mathcal{L}^{wEH}= S + \frac{\alpha}{90\pi} \frac{1}{B_{\rm cr}^2}\left(4S^2+7P^2\right)  + \mathcal O(\alpha^2),
    \label{eq:wEH}
\end{equation}

\begin{equation}
    \mathcal{L}^{sEH}= S - \frac{\alpha}{3\pi}S\left(\ln\frac{E_{\rm cr}}{\sqrt{-2S}}+\kappa\right) + \mathcal O(\alpha^2),
    \label{eq:sEH}
\end{equation}

where $\alpha$ is the fine structure constant, $\kappa = \ln\pi +\gamma -6\zeta'(2)/\pi^3$, $\tilde F_{\mu\nu}$ represents the Hodge dual of the electromagnetic tensor.\\

Most work in the recent literature focuses on the weak Euler - Heisenberg theory described by (\ref{eq:wEH}), leaving sEH theory barely explored. We extended that work to analyze light propagation and Cherenkov radiation in magnetic fields above Schwinger's and compared them with the weaker scenario.\\

In a recent paper, Macleod et al \cite{Macleod2019}, analyze the properties of the vacuum Cherenkov radiation in laser pulses and the magnetic field around a pulsar, finding regimes in which it is the dominant radiation mechanism, compared with the synchrotron radiation, which is also emitted by the charge under the influence of the magnetic field. This radiation process may be relevant to the excess signals of high energy photons in astrophysical observations \cite{Kaspi2017}.\\

It has been questioned  \cite{Lee2020} that Cherenkov radiation eventually would dominate the synchrotron radiation because in  effective field theory,  the maximal attainable photon frequency $\omega_{\rm max}$ is about four orders of magnitude lower than the critical synchrotron frequency $\omega_{\rm cr}$.
This situation would change for supercritical magnetic fields, that are up to 2 orders of magnitude above Schwinger's, since the cut-off frequency and the radiated power depend on the applied magnetic field.\\

The strong-Euler-Heisenberg theory applies for supercritical fields of order $B \sim 10^2\,B_{\rm cr} \approx 10^{11}\,T$, while weak-Euler-Heisenberg is appropriate for fields of the order of the Schwinger field or below, $B_{\rm cr} \approx 10^{9}T$. Since the sEH field is two orders of magnitude greater than the Schwinger's critical field, then the properties of vacuum in the presence of the sEH field become more auspicious for the Cherenkov radiation to be observed in astrophysical and extreme environments. Particularly, magnetars have magnetic fields of order $\sim 10^{11}T$ \cite{Kaspi2017}, thus providing a suitable ground to astrophysical measurements of non-linear electromagnetic effects.\\

In this work we concentrate in studying the effects of sEH in a flat background as a first approximation to the problem. Of course, a more realistic scenario would be to study this phenomenon in a curved spacetime such as the charged Euler - Heisenberg black hole \cite{Ruffini2013}; a magnetized black hole described by the Ernst metric \cite{Ernst1976}, that consists in a Schwarzschild black hole metric embedded in an external magnetic field, or a magnetized rotating black hole which was found in \cite{Breton1986}.\\
 
We determine the Cherenkov angle and the bounds on the charge velocity for the occurrence of the Cherenkov radiation according to the  strong-Euler-Heisenberg (sEH) theory.   Considering supercritical magnetic fields, the refractive index of the resulting sEH environment is larger than the one due to the wEH vacuum polarization and the photon phase velocities are slower than the ones predicted by wEH. Then the Cherenkov effect can be achieved with slower charged particles, whose synchrotron radiation is also smaller, as well as the maximum frequency.\\

The paper is organized as follows:\\

In Section \ref{section2} it is discussed how the photon trajectories are described in nonlinear electrodynamics (NLED) by the null geodesics of an optical metric. Then we derive the phase velocity and refractive index that result from the NLED-light interaction in the sEH theory for a Minkowskian background metric and use it to compare the diminishment in light velocity produced by the sEH and wEH theories for different magnetic field intensities.\\

In Section \ref{section3} we study the conditions under which Cherenkov radiation may occur. We then define the Cherenkov angle in terms of quantities obtained in \ref{section2} so it directly ties to NLED and geometric parameters. Finally, in this section we provide computations for the speed velocity and refractive index for several magnetic field both in the wEH and the sEH theories. \\

In Section \ref{section4} we compare the synchrotron radiation with the Cherenkov radiation produced by  sEH and wEH  NLEDs. We considered different values of magnetic field and a fixed Lorentz $\gamma$ factor for the travelling particle. The presence of birefringence in wEH theory yields two different threshold velocities for Cherenkov radiation to be emitted. We also studied how the crossing frequency, above which Cherenkov radiation dominates over synchrotron, varies with the magnetic field intensity.\\

Conclusions and future work are presented in Section \ref{section5}. We confirmed our guess that for supercritical fields in the sEH regime the excited frequencies are in the radiofrequencies range, opening a new perspective for observations in the low frequencies band. Therefore, sEH radiation provides a better chance to be observed than the expectations coming from the wEH theory, as long as the background field meets the conditions to be described by the sEH theory.\\

Through this paper, unless it is explicitly stated otherwise, we will work with geometric units $\hbar = c = G = 1$ and we will use the {\it mostly pluses} convention for the metric tensor.


\section{The effective or optical metric in nonlinear electrodynamics.}
\label{section2}

It is well known that intense electromagnetic (EM) fields, where the Maxwell theory is no longer valid, can resemble a curved spacetime, in the sense that light trajectories are not straight lines but undergo deflection. Deviations from the straight trajectories in vacuum are described in NLED by the null trajectories of an effective or optical metric. The effective metric is obtained from the analysis of the propagation of discontinuities or perturbations of the EM field, \cite{Novello2000b} that is, according to this formalism the EM fields of the propagating wave are much smaller than the background fields. The effective optical metric approach turns out to be equivalent to the soft photon approximation \cite{Novello2000b}. Considering that $k_{\mu}$ is a vector normal to the characteristic surface of the wave, the effective optical metric $\gamma_{\mu \nu}$ determines the dispersion relation of the photon wave as:

\begin{equation}
    \gamma^{\mu \nu}_{(i)} k_{\mu} k_{\nu} = 0, \quad i= \pm,
    \label{effmetr}
\end{equation}

where the $(i)$ subscript corresponds to the two metrics that can arise in NLED, when the phenomenon of birefringence occurs, so the polarization of the photon with respect to the field determines which metric it ``sees". For a light ray propagating along  with wave vector $k^{\mu} =( \omega, k^{x}, k^{y}, k^{z})$, its phase velocity is obtained from Eq.(\ref{effmetr}), that is the dispersion relation, as

\begin{equation}
    \gamma_{ t t} \omega^2 + 2 \gamma_{i t} \omega k^i + \gamma_{i j }k^{i}k^{j} = 0.
    \label{disp_eq}
\end{equation}

where latin indices indicate the spatial direction, $i,j = x,y,z$. The phase velocity of light is defined in terms of the wave vector as the quotient $v_{\rm ph}=\omega/k$, being $k$ the magnitude of the spatial part calculated with the background metric as $k^2 = g_{ij}k^ik^j$; then the phase velocity is given by

\begin{equation}
    v_{\rm ph}= \frac{1}{\gamma_{ t t} k} \left\{ -{\gamma_{ i t}}k^i \pm  \sqrt{  \gamma_{i t}\gamma_{jt}k^ik^j  -\gamma_{t t } \gamma_{i j} k^{i}k^{j}}  \right\}.
    \label{vph}
\end{equation}

The effective metrics can be computed for a given, generic, NLED Lagrangian $\mathcal{L}(S,P)$ as
\begin{align}
    \gamma^{\mu\nu}_{(-)}& = \left(\mathcal{L}_S+\frac12 \mathcal{L}_{PP} S\right)g^{\mu\nu} + \mathcal{L}_{PP}F\indices{^\mu_\lambda}F^{\lambda\nu},\label{eq:gamma1}\\  
    \gamma^{\mu\nu}_{(+)}& = \mathcal{L}_{S} g^{\mu\nu} + \mathcal{L}_{SS}F\indices{^\mu_\lambda}F^{\lambda\nu},\label{eq:gamma2} 
\end{align}

where $g_{\mu\nu}$ is the background metric, that in absence of gravity is the Minkowski space, $\eta_{\mu\nu}= {\rm diag} [-1,1,1,1]$; we note that the above expressions for the effective metric are valid in the case $\mathcal{L}_{SP}=0$. The ``--" and ``+" subscripts indicate that the electric field of the propagating linearly polarized light is parallel or orthogonal to the magnetic field lines, respectively.\\

Then, to determine the EH phase velocity from (\ref{vph}) we need to know the background metric $g_{\mu \nu}$ and electromagnetic field  $F_{\mu \nu}$ in the specific system; for a diagonal background metric it turns out that,
\begin{equation}
    v_{\rm ph}= \sqrt{\frac{\gamma_{i j}}{- \gamma_{t t}} \left( \frac{k^{i}k^{j}}{k^2} \right)},
\end{equation} 

We focus on the pure magnetic case, therefore the invariant $P = -\vec E\cdot\vec B$ identically vanishes. In the strong magnetic field scenario this will imply that, according to Eq.(\ref{eq:sEH}), one of the effective metrics, Eq.(\ref{eq:gamma1}), is conformal to the background metric, and therefore photon trajectories remain unaffected. This will mean that only one polarization mode will feel the NLED effects. On the other hand, Eq.(\ref{eq:gamma1}) is non-trivial for the wEH scenario and trajectories will split into two effective cones, none of which coincides with the background one. We will refer to them as ``+ modes" and ``- modes".\\

That being said, we consider a constant and spatially uniform magnetic field, $B_0$, in the Cartesian $x$-direction with no electric field. Therefore, the electromagnetic field tensor and the $S$ invariant are given, respectively, by

\begin{equation}
    F_{\mu\nu} = 2 B_0 \delta_{[\mu}^z \delta_{\nu]}^y;\qquad S = - \frac{1}{2}B_0^2.
    \label{eq:emtensor}
\end{equation}
 
Also, since the dispersion relation is given by $\gamma^{\mu\nu} k_\mu k_\nu =0$, we can divide the expressions Eq.(\ref{eq:gamma1}) and Eq.(\ref{eq:gamma2}) by $\mathcal L_S + S\mathcal L_{PP}/2$ and $\mathcal{L}_S$ respectively and work with an equivalent yet simpler effective metric. This form also makes clearer what is the deviation from the background metric and allows to better determine the NLED effects. The effective line elements are then given by

\begin{align}
    d\sigma^2_- &= \gamma^{(-)}_{\mu\nu}dx^\mu dx^\nu = -dt^2 +  dx^2 + \left(1 + \frac{2 \mathcal L_{PP}}{2 \mathcal L_{S} + S\mathcal L_{PP}} B_0^2\right) (dy^2+dz^2), \\
    d\sigma^2_+ &= \gamma^{(+)}_{\mu\nu}dx^\mu dx^\nu = -dt^2 +  dx^2 + \left(1 + \frac{\mathcal L_{SS}}{\mathcal L_{S}} B_0^2\right) (dy^2+dz^2),
\end{align}

which particularizes for the sEH and wEH theories as

\begin{align}
    d\sigma^2(wEH^-) &= -dt^2 +  dx^2 +  \left(1 - \frac{7\alpha}{45\pi} \frac{B_0^2}{B_{\rm cr}^2}\right) (dy^2+dz^2) + \mathcal O(\alpha^2),\label{eq:wehg-} \\
    d\sigma^2(wEH^+) &= -dt^2 +  dx^2 + \left(1 - \frac{4\alpha}{45} \frac{B_0^2}{B_{\rm cr}^2}\right) (dy^2+dz^2) + \mathcal O(\alpha^2),\label{eq:wehg+} \\
    d\sigma^2(sEH)& = -dt^2 + dx^2 + \left(1 + \frac{\alpha}{6\pi} \frac{B_0^2}{S}\right) (dy^2+dz^2) + \mathcal O(\alpha^2).
    \label{eq:sehg}
\end{align}

Where for the sEH effective metric we have omitted the superscript "+" as there is only one optical metric and also note that in the pure magnetic case $S<0$; therefore, the correction does indeed slow down light. We shall consider that the $S$ invariant involves fields of the order of the critical magnetic field.

\section{Conditions for the Cherenkov radiation.}
\label{section3}

Cherenkov radiation will occur whenever a charged particle travels through a region of space (typically a dielectric material, but not necessarily) with a speed above light phase velocity in that region. Equations (\ref{eq:wehg+}), (\ref{eq:wehg-}) and (\ref{eq:sehg}) together with expression (\ref{vph}) allow us to calculate the correction to the phase velocity of light due to the presence of the strong magnetic field.
We can then express the phase velocity as $v_{\rm ph} = 1 + \delta v_{\rm ph}$ where the correction is $\mathcal O(\alpha)$. These corrections lead to the following phase velocities for the wEH and sEH theories, respectively,
\begin{align}
    v_{\rm ph}^{wEH-}& = \sqrt{1 - \frac{7\alpha}{45\pi} \frac{B_0^2}{B_{\rm cr}^2}} = 1 - \frac{7\alpha}{90\pi} \frac{B_0^2}{B_{\rm cr}^2} + \mathcal O(\alpha^2), \\
    v_{\rm ph}^{wEH+} &= \sqrt{1 - \frac{4\alpha}{45\pi} \frac{B_0^2}{B_{\rm cr}^2}} = 1 - \frac{2\alpha}{45\pi} \frac{B_0^2}{B_{\rm cr}^2} + \mathcal O(\alpha^2), \\
    v_{\rm ph}^{sEH} &= \sqrt{1{-} \frac{\alpha}{6\pi} \frac{B_0^2}{{B_{\rm cr}^2}}} = 1 {-} \frac{\alpha}{12\pi} \frac{B_0^2}{{B_{\rm cr}^2}} + \mathcal O(\alpha^2).
\end{align}

These velocities set the threshold for Cherenkov radiation to occur since the charge must be faster than the phase velocity of light in the polarized vacuum. Thus, in order to produce Cherenkov radiation, a charged particle needs to reach a speed
\begin{align}
    \beta^{wEH-} &> 1 - \frac{7\alpha}{90\pi} \frac{B_0^2}{B_{\rm cr}^2} + \mathcal O(\alpha^2), \label{eq:beta_min1}\\
    \beta^{wEH+} &> 1 - \frac{2\alpha}{45\pi} \frac{B_0^2}{B_{\rm cr}^2} + \mathcal O(\alpha^2), \label{eq:beta_min2}\\
    \beta^{sEH} &> 1 - \frac{\alpha}{12\pi} \frac{B_0^2}{{B_{\rm cr}^2}} + \mathcal O(\alpha^2).
    \label{eq:beta_min3}
\end{align}

 Comparing the correction of the phase velocity for both wEH polarization modes, we find that $\delta v_{\rm ph}^{wEH-} / \delta v_{\rm ph}^{wEH+} \approx 2$, implying that the threshold for Cherenkov radiation to occur is lower in the parallel polarization case. Consequently, wEH birefringence favors the emission of Cherenkov photons polarized parallel to the magnetic field, which are the main contribution to the total radiated power (this can also be seen in Fig.(\ref{fig:wEH-Sync})).

\begin{table}[htbp]
\centering
    \begin{tabular}{c|cccc}
       $B [B_{\rm cr}]$\,\, & \,\,$v_{ph}^{+}$ & $v_{ph}^{-}$ & $n^+$ & $n^-$  \\ \hline
        0.1  & 0.999998 & 0.999998  & 1.000001 & 1.000001 \\
        0.3  & 0.999990 & 0.999983 & 1.000009  & 1.000016 \\
        0.5  & 0.999974 & 0.999954 & 1.000025 & 1.000045 \\
        0.7  & 0.999949 & 0.999911 & 1.000050 & 1.000088 \\
        1    & 0.999875 & 0.999781 & 1.000124 & 1.000218 \\
        2    & 0.999544 & 0.999202 & 1.000455 & 1.000796 \\
        3    & 0.999007 & 0.998261 & 1.000992 & 1.001736 \\
        4    & 0.998428 & 0.997247 & 1.001570 & 1.002748 \\
        5    & 0.997310 & 0.995288 & 1.002685 & 1.004700 \\
        6    & 0.996398 & 0.993689 & 1.003594 & 1.006290 \\
        7    & 0.994780 & 0.990848 & 1.005205 & 1.009109 \\
        8    & 0.993201 & 0.988072 & 1.006775 & 1.011856 \\
        9    & 0.991786 & 0.985581 & 1.008179 & 1.014314 \\
        10   & 0.989410 & 0.981392 & 1.010533 & 1.018434
    \end{tabular}
    \caption{Phase velocity and refractive index for both + and - polarization modes in the wEH theory for different values of the magnetic field around the Schwinger field. For lower magnetic fields, although present, the corrections are much smaller.}
    \label{tab:wEH}
\end{table}

\begin{table}[htbp]
\centering
    \begin{tabular}{c|cc}
        $B [B_{\rm cr}]$\,\, & \,\,$v_{ph}^+$ & $n^+$  \\ \hline
        5  & 0.99516 & 1.00486  \\
        10 & 0.98064 & 1.01974  \\
        15 & 0.95644 & 1.04555  \\
        20 & 0.92255 & 1.08395  \\
        25 & 0.87899 & 1.13767  \\
        30 & 0.82574 & 1.21103  \\
        35 & 0.76282 & 1.31093  \\
    \end{tabular}
    \caption{Phase velocity and refractive index for the + polarization mode in the sEH theory. It can be seen that $B_0 = 35B_{\rm cr}$ already yields a refraction index of water. For stronger magnetic fields the $\mathcal O(\alpha)$ approximation is not enough to guarantee trustable results.}
    \label{tab:sEH}
\end{table}

It is instructive to think of the Cherenkov angle in terms of the induced refractive index of vacuum, so comparisons with dielectric materials become clearer. This also illustrates the magnetic field required to simulate common results in laboratory optics and how their magnitude relates; for instance, to achieve the refractive index of water, one would need a magnetic field of order of $B_0 \approx 35 B_{\rm cr}$.


\begin{figure}[htbp]
    \centering
    \includegraphics[width = 0.5\textwidth]{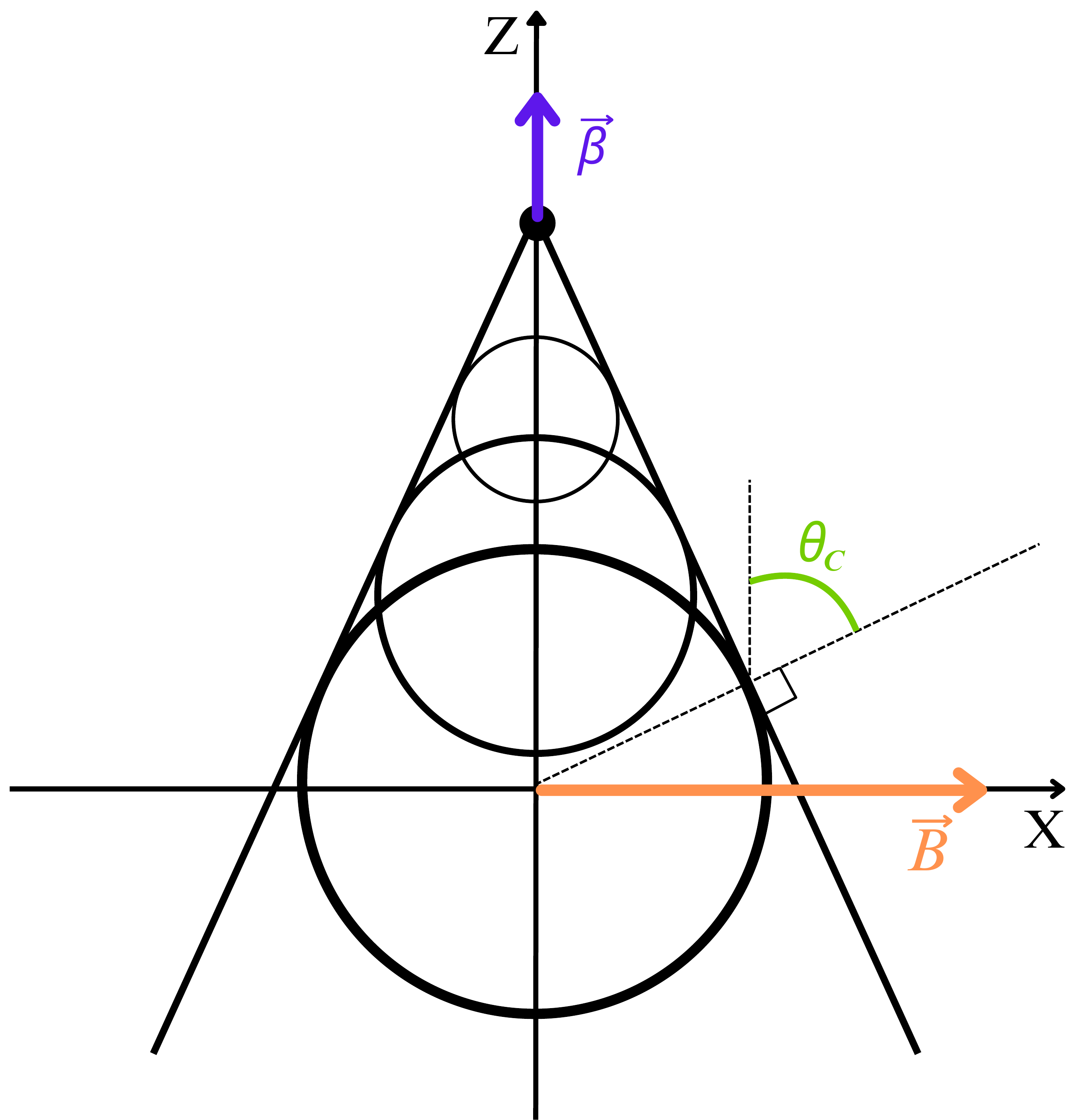}
    \caption{Cherenkov cone shown for a particle travelling in the $z$-direction with velocity $\vec\beta$ (blue arrow) perpendicular to an external magnetic field $\vec B$ (orange arrow). The circles represent the wavefronts emitted by the particle and the Cherenkov angle is shown in green.}
    \label{fig:cone}
\end{figure}

Whenever Cherenkov radiation occurs, it can be parametrized with a cone defined as the tangent surface to the wavefront left behind by the charged particle and the instantaneous position of the particle itself, called the Cherenkov cone. This geometric setup is shown in figure Fig.~(\ref{fig:cone}). Such a cone is characterized by its internal angle, defined by \cite{Jackson}

\begin{equation}
    \cos^2 \theta_{C}= \frac{v_{\rm ph}^2}{\beta^2},
\end{equation}
where  $v_{\rm ph}$ is the phase velocity of light in a given polarization mode and $\beta$ is the charge's velocity; the condition for Cherenkov radiation to occur is $\cos^2 \theta_{C} <1$. We can associate a refractive index with the polarized vacuum as the quotient between the speed of light in the absence of electromagnetic fields and the modified speed, $n(x) = 1/v_{ph}$. The refraction index can also depend on the frequency, but this is beyond the scope of this work. In terms of this refractive index, the Cherenkov angle is thus defined through

\begin{equation}
    \cos^2 \theta_C = \frac{1}{n^2 \beta^2}.
\end{equation}

Tables \ref{tab:wEH} and \ref{tab:sEH} show the phase velocity of light in the polarized vacuum for different values of the magnetic field, together with the refractive index induced in the vacuum for the wEH and sEH theories, respectively. Interestingly, the phase velocity and refraction index for magnetic fields of medium intensity, like $B=5B_{\rm cr}$ or $B=10B_{\rm cr}$, calculated in wEH and sEH differ by about 1\%, suggesting a smooth transition from one theory to the other.


\section{Cherenkov and synchrotron Radiated powers.}
\label{section4}

Since the purpose of this section is to compare the emitted frequencies and provide a reference for their intensity. Through this part,  we will use $\hbar = 6.582\times10^{-16}$eV$\cdot$s and keep $c=1$. The electron mass is taken to be $m_e = 511$keV and its charge is obtained from the fine structure constant through $e^2 = 4\pi\hbar\alpha$.\\

In the presence of a strong magnetic field and with the electrons satisfying the stated condition Eq.(\ref{eq:beta_min1})-(\ref{eq:beta_min2})-(\ref{eq:beta_min3}), they will emit both Cherenkov and synchrotron radiation.\\


MacLeod et al \cite{Macleod2019} obtained the expression for the radiated Cherenkov power per frequency unit within a non-linear electromagnetic, non-isotropic background. Since the Euler-Heisenberg theory produces vacuum birefringence, the emitted power will be dependent on photon polarization,

\begin{equation}
    \frac{d^2P_{Cher}^\pm}{d\omega d\phi} = \frac{\alpha}{2\pi}\omega |\hat \epsilon_0\cdot\hat\epsilon_\pm|^2\sin^2\theta_c^\pm,
    \label{eq:cherenkov}
\end{equation}

where $\pm$ indicates which polarization, i.e. which light phase speed we are considering; and the unit vectors $\hat\epsilon_\pm$ and $\hat\epsilon_0$ are the unit spatial components of the polarization modes and the polarization of the Cherenkov photons. Non-isotropic behavior of polarized vacuum induces birefringence, which implies that only the projection along the directions of $\hat\epsilon_\pm$ will be emitted, decreasing the total Cherenkov power emitted with respect to a perfectly isotropic and homogeneous medium. From Eq.(\ref{eq:cherenkov}), it is straightforward to obtain the complete Cherenkov spectrum by integrating over the azimuthal angle and adding both contributions:

\begin{equation}
    \frac{dP_{Cher}}{d\omega} = \int_0^{2\pi} \left(\frac{d^2P_{Cher}^+}{d\omega d\phi} + \frac{d^2P_{Cher}^-}{d\omega d\phi}\right)d\phi
    \label{eq:cherenkovtotal}
\end{equation}

Each of the polarization amplitudes was calculated by Macleod \cite{Macleod2019} and are given by

\begin{align}
    |\hat \epsilon_0\cdot\hat\epsilon_+|^2 &= \frac{\cos^2\phi}{1-\sin^2\theta_c^+\sin^2\phi},\label{eq:amplitud}\\
    |\hat \epsilon_0\cdot\hat\epsilon_-|^2 &= \frac{\sin^2\phi\cos^2\theta_c^-}{1-\sin^2\theta_c^-\sin^2\phi}.
    \label{eq:amplitudes}
\end{align}


Equations Eq.(\ref{eq:amplitudes})-(\ref{eq:amplitud}) are valid when the particle travels in the positive {\it z}-direction and the Cherenkov angle is aligned with the usual $\theta$ from spherical coordinates, as it is shown in Fig. \ref{fig:cone}. On the other hand, the expression for the synchrotron radiation was first computed by Schwinger \cite{Schwinger}:

\begin{equation}
    \frac{dP_{Sync}}{d\omega} = \frac{\sqrt 3}{\pi}\frac{e^3}{m_e}B_0\frac{\omega}{\omega_c}\int_{\omega/\omega_c}^\infty K_{5/3}(x)dx
    \label{eq:sync}
\end{equation}

Where $e$ is the elementary charge, $\gamma$ is the Lorentz factor of the particle, $K_{5/3}(x)$ is the modified Bessel function, and $\omega_c = 3\gamma^3 B_0e/(2 m_e)$ is the characteristic frequency of the system that depends on the field generating the non-linearities. Also, whenever the external field is expressed in units of the critical field as $B_0 = \lambda B_{\rm cr}$, we can insert this into the expression for the characteristic frequency to get $\omega_c \equiv \omega_\lambda = \lambda (3\gamma^3 m_e/2)$ and rewrite Eq.(\ref{eq:sync}) in a simpler fashion as

\begin{equation}
    \frac{dP_{Sync}}{d\omega} = 4\sqrt 3\, \alpha m_e\frac{\omega}{\omega_1}\int_{\omega/\omega_\lambda}^\infty K_{5/3}(x)dx,
\end{equation}

where a Lorentz factor of $\gamma=10^5$ yields $\omega_1 \approx 1.533 \times10^{21}$s$^{-1}$. Lee \cite{Lee2020} examined the range of synchrotron vs Cherenkov radiation and suggested that the Cherenkov power will dominate over synchrotron at frequencies way below $\omega_c$; therefore, we can use the low-frequency approximation of Eq.(\ref{eq:sync}) to simplify the calculations. This approximation is given by \cite{Jackson} 

\begin{equation}
    \frac{dP_{Sync}}{d\omega} = 2\,3^{1/3}\frac{\Gamma(2/3)^2}{\pi}\alpha\gamma^2\left(\frac{\omega}{\omega_c}\right)^{1/3}
    \label{eq:lowsync}
\end{equation}

\begin{figure}[htbp]
    \centering
    \includegraphics[width = \textwidth]{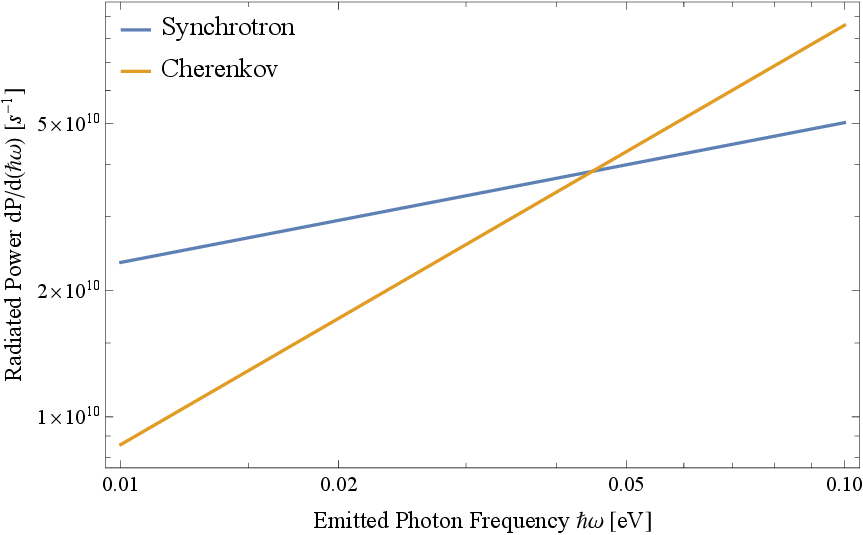}
    \caption{Radiated Cherenkov power due to an electron travelling with a Lorentz factor of $\gamma = 10^5$ through a magnetic field $30 B_{\rm cr}$ in the regime of the sEH.}
    \label{fig:sEH-Sync}
\end{figure}

\begin{figure}[htbp]
    \centering
    \includegraphics[width = \textwidth]{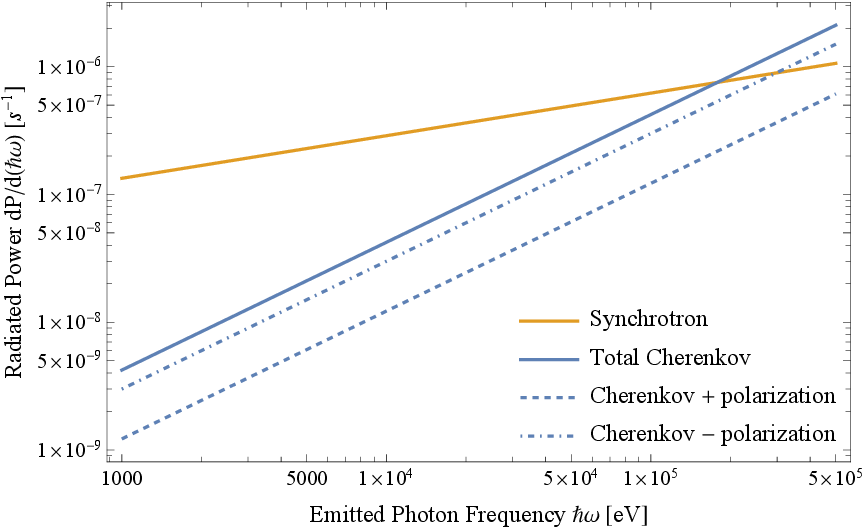}
    \caption{Radiated Cherenkov power due to an electron travelling with a Lorentz factor of $\gamma = 10^5$ through a magnetic field equal to $10^{-3}B_{\rm cr}$. For this case the weak field limit is considered. The dashed and dot-dashed lines represent the contributions of each polarization to the total Cherenkov power. In accordance with table \ref{tab:wEH} the - mode radiates higher power than the + mode since the threshold velocity is lower.}
    \label{fig:wEH-Sync}
\end{figure}

\begin{figure}[htbp]
\centering
    \includegraphics[width = \textwidth]{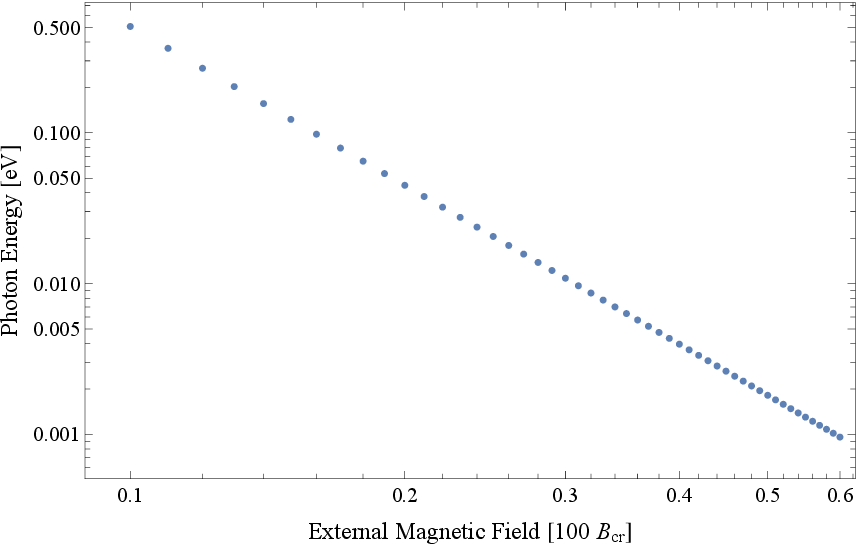}
    \caption{Emitted photon energy for which the Cherenkov radiation and the synchrotron emission coincide for a given magnetic field in the sEH. The critical field has been rescaled by a factor of 100 to avoid machine-precision issues.}
    \label{fig:sEH=Sync}
\end{figure}

\begin{figure}[htbp]
\centering
    \includegraphics[width = \textwidth]{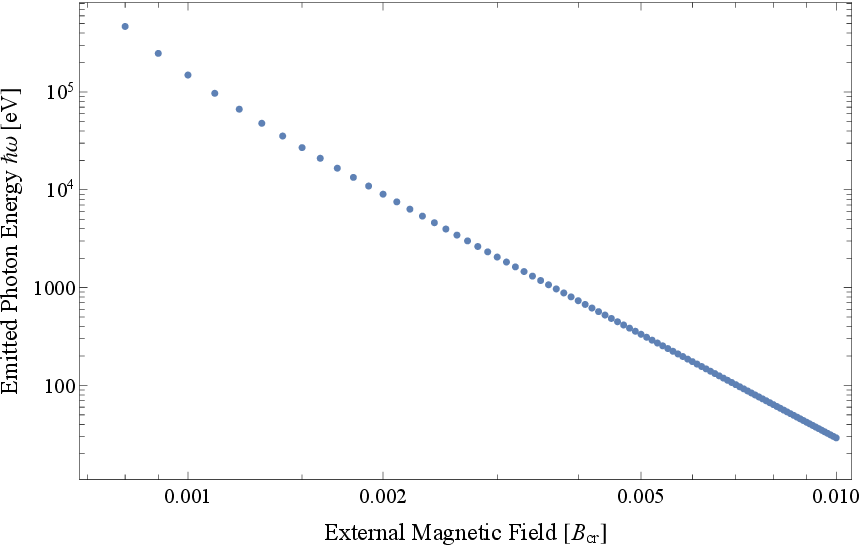}
    \caption{Emitted photon energy for which the Cherenkov radiation and the synchrotron emission coincide for a given magnetic field in the wEH.}
    \label{fig:wEH=Sync}
\end{figure}

Now we can compare how each radiation contributes to the total power measured. The radiated synchrotron and Cherenkov power spectrum is shown in Fig. \ref{fig:sEH-Sync} and Fig. \ref{fig:wEH-Sync}. From these two figures, it can be directly seen how, in fact, sEH Cherenkov radiation dominates over synchrotron for frequencies $\omega\sim 10^8-10^9 s^{-1}$; in contrast, in wEH the dominance starts at frequencies of order $\omega \sim 10^{23}s^{-1}$. Stronger fields result in a more dielectric-like behavior of the vacuum; therefore, the Cherenkov response of the vacuum clearly intensifies. This behavior occurs both in the strong and weak field limits, as can be seen in  Fig. \ref{fig:sEH=Sync} and Fig. \ref{fig:wEH=Sync}, where are shown  the emitted photon energy for which the Cherenkov radiation and the synchrotron emission coincide for a given magnetic field in the sEH and wEH, respectively.

\section{Conclusions and future work.}\label{section5}

In this work we have analyzed the conditions for Cherenkov radiation to occur in vacuum in the presence of critical and supercritical magnetic field of the Euler - Heisenberg non-linear electrodynamics. We first derived the effective metrics that arise in the presence of NLED that define the dispersion relations satisfied by light waves; these metrics account for the vacuum polarization effects, making it mimic the structure of a dielectric medium. We noted that to get an effective refractive index of the order of common materials, like water, the magnetic field needs to reach an intensity of $35 B_{\rm cr} \sim 10^{12}\,T$ in the regime of strong magnetic fields. This is near the typical magnetic field of a magnetar \cite{Kaspi2017}; therefore, astrophysical compact objects are the most suitable systems for NLED effects to be observed.\\ 

Particularly, for both the weak and strong EH field approximations, we determine the phase velocity of light in vacuum as a function of the external critical magnetic field responsible for the vacuum polarization. In general, this decreasing of the speed depends upon the polarization of the propagating wave with respect to the field lines; however, for our case of study, this birefringence phenomenon is only observed in the weak field approximation. Thanks to this reduction in the speed of light, Cherenkov radiation may occur in vacuum when a charged particle travels faster than the given phase velocity. We then took advantage of this fact to calculate the threshold velocities of such a particle to emit Cherenkov radiation as a function of the magnetic field. \\

We found that, within their respective range of validity, the strong-EH systematically requires a lower speed than the weak-EH to produce Cherenkov radiation, as expected. We studied this phenomenon for magnetic fields measured as multiples of the critical Schwinger field. Additionally, we proved that the lower bound for Cherenkov radiation in the weak magnetic field scenario is given by one of the polarizations, whose related threshold speed remains lower than the other regardless of the intensity of the magnetic field. Cherenkov radiation only requires the charged particle to travel faster than light, independently of the polarization state, meaning that it will be emitted as soon as the particle exceeds one of the speeds of light. Also, for a speed between the two thresholds, the emitted radiation will necessarily be polarized parallel to the cone whose speed was exceeded, as can be inferred from Eq.(\ref{eq:cherenkov}).\\

Then, since affected by an external field a charged particle will emit synchrotron radiation \cite{Schwinger}, we analyze the conditions to be satisfied so that Cherenkov radiation dominates over synchrotron. In particular, we determine which frequency marks that point. We found that in the case of strong fields (sEH) the intersection of the two emitted powers occurs at  a lower frequency than the one for the wEH. In addition, we noted that the threshold frequency decreases as the magnetic field is increased, in agreement with the result we just stated. 

Although promising, these results are yet a first step towards the astrophysical problem. In future work we expect to extend the calculations to curved backgrounds that describe compact objects. It is clear that a deeper analysis will boost the chances of getting testable astrophysical predictions. As for now, we expect that introducing a curved metric $g_{\mu\nu}$ in (\ref{effmetr}) will have a direct impact in most our calculations. First, there would appear closed lightlike curves; second, the magnetic field would be given a different, more precise, shape, affecting the behaviour of differently polarized light rays. Also, the presence of causal horizons might be a relevant constrain to the proportion of emitted radiation we can measure. Since sEH theory is barely explored, we expect fruitful and extense research in this area in the coming years.
\acknowledgments
NB acknowledges partial support of SECIHTI-Mexico
project CBF2023-2024-811. The work of DGG has been sponsored by Conahcyt-Mexico (SECIHTI-Mexico) through the Ph.D. scholarship No.4046418. DGG also wants to acknowledge his closest family members for their unconditional support, especially his beloved mother Antonia; as well as his colleagues and friends who accompanied him during this journey.

\bibliographystyle{JHEP}
\bibliography{biblio.bib}

@article{BI,
    author = "Born, M. and Infeld, L.",
    title = "{Foundations of the new field theory}",
    doi = "10.1098/rspa.1934.0059",
    journal = "Proc. Roy. Soc. Lond. A",
    volume = "144",
    number = "852",
    pages = "425--451",
    year = "1934"
}

@article{EH,
      title={Consequences of Dirac Theory of the Positron}, 
      author={W. Heisenberg and H. Euler},
      year={2006},
      eprint={physics/0605038},
      archivePrefix={arXiv},
      primaryClass={physics.hist-ph},
      url={https://arxiv.org/abs/physics/0605038}
}

@article{Aiello2007,
    title = {Anisotropic effects of background fields on Born–Infeld electromagnetic waves},
    journal = {Physics Letters A},
    volume = {361},
    number = {1},
    pages = {9-12},
    year = {2007},
    issn = {0375-9601},
    doi = {https://doi.org/10.1016/j.physleta.2006.09.027},
    url = {https://www.sciencedirect.com/science/article/pii/S0375960106014289},
    author = {Matías Aiello and Gabriel R. Bengochea and Rafael Ferraro}
}

@article{Ruffini2013,
  title = {Einstein-Euler-Heisenberg theory and charged black holes},
  author = {Ruffini, Remo and Wu, Yuan-Bin and Xue, She-Sheng},
  journal = {Phys. Rev. D},
  volume = {88},
  issue = {8},
  pages = {085004},
  numpages = {13},
  year = {2013},
  month = {Oct},
  publisher = {American Physical Society},
  doi = {10.1103/PhysRevD.88.085004},
  url = {https://link.aps.org/doi/10.1103/PhysRevD.88.085004}
}

@article{Macleod2019,
  title = {Cherenkov Radiation from the Quantum Vacuum},
  author = {Macleod, Alexander J. and Noble, Adam and Jaroszynski, Dino A.},
  journal = {Phys. Rev. Lett.},
  volume = {122},
  issue = {16},
  pages = {161601},
  numpages = {6},
  year = {2019},
  month = {Apr},
  publisher = {American Physical Society},
  doi = {10.1103/PhysRevLett.122.161601},
  url = {https://link.aps.org/doi/10.1103/PhysRevLett.122.161601}
}

@article{Kaspi2017,
   author = "Kaspi, Victoria M. and Beloborodov, Andrei M.",
   title = "Magnetars", 
   journal= "Annual Review of Astronomy and Astrophysics",
   year = "2017",
   volume = "55",
   number = "Volume 55, 2017",
   pages = "261-301",
   doi = "https://doi.org/10.1146/annurev-astro-081915-023329",
   url = "https://www.annualreviews.org/content/journals/10.1146/annurev-astro-081915-023329",
   publisher = "Annual Reviews",
   issn = "1545-4282",
   type = "Journal Article"
}

@article{Lee2020,
    title = {Cherenkov radiation in a strong magnetic field},
    journal = {Physics Letters B},
    volume = {810},
    pages = {135794},
    year = {2020},
    issn = {0370-2693},
    doi = {https://doi.org/10.1016/j.physletb.2020.135794},
    url = {https://www.sciencedirect.com/science/article/pii/S0370269320305979},
    author = {Cheng-Yang Lee}
}

@article{Karbstein2019,
  title = {All-Loop Result for the Strong Magnetic Field Limit of the Heisenberg-Euler Effective Lagrangian},
  author = {Karbstein, Felix},
  journal = {Phys. Rev. Lett.},
  volume = {122},
  issue = {21},
  pages = {211602},
  numpages = {5},
  year = {2019},
  month = {May},
  publisher = {American Physical Society},
  doi = {10.1103/PhysRevLett.122.211602},
  url = {https://link.aps.org/doi/10.1103/PhysRevLett.122.211602}
}

@article{Schwinger1951,
    title={On Gauge Invariance and Vacuum Polarization}, 
    volume={82}, 
    url={http://dx.doi.org/10.1103/physrev.82.664}, 
    DOI={10.1103/physrev.82.664}, 
    number={5}, 
    journal={Physical Review}, 
    publisher={American Physical Society (APS)},
    author={Schwinger, Julian}, 
    year={1951}, 
    month={june}, 
    pages={664–679}
}

@article{Novello2000b,
   title={Light propagation in non-linear electrodynamics},
   volume={482},
   ISSN={0370-2693},
   url={http://dx.doi.org/10.1016/S0370-2693(00)00522-0},
   DOI={10.1016/s0370-2693(00)00522-0},
   number={1-3},
   journal={Physics Letters B},
   publisher={Elsevier BV},
   author={De Lorenci, V.A. and Klippert, R. and Novello, M. and Salim, J.M.},
   year={2000},
   month={June},
   pages={134–140}
}

@book{Jackson,
    author = "Jackson, John David",
    title = "{Classical Electrodynamics}",
    isbn = "978-0-471-30932-1",
    publisher = "Wiley",
    year = "1998"
}

@article{Schwinger, 
    title={On the Classical Radiation of Accelerated Electrons}, 
    volume={75}, 
    url={http://dx.doi.org/10.1103/physrev.75.1912},
    DOI={10.1103/physrev.75.1912},
    number={12},
    journal={Physical Review},
    publisher={American Physical Society (APS)},
    author={Schwinger, Julian}, 
    year={1949},
    month={june}, 
    pages={1912–1925}
}

@article{Ernst1976,
       author = {Ernst, F.~J.},
        title = {Black holes in a magnetic universe},
      journal = {Journal of Mathematical Physics},
         year = {1976},
        month = {jan},
       volume = {17},
       number = {1},
        pages = {54-56},
          doi = {10.1063/1.522781},
       adsurl = {https://ui.adsabs.harvard.edu/abs/1976JMP....17...54E}
}

@article{Breton1986,
    author = {Bretón Baez, Nora and García Díaz, Alberto},
    title = {The most general magnetized Kerr–Newman metric},
    journal = {Journal of Mathematical Physics},
    volume = {27},
    number = {2},
    pages = {562-563},
    year = {1986},
    month = {02},
    issn = {0022-2488},
    doi = {10.1063/1.527208},
    url = {https://doi.org/10.1063/1.527208}
}

\end{document}